\documentclass[reprint,aps,prl,nobibnotes,twocolumn]{revtex4-2} 

\usepackage{amsmath}
\usepackage{amssymb}
\usepackage{graphicx}
\usepackage{fullpage}
\usepackage{hyperref}

\usepackage{color}
\definecolor{darkblue}{rgb}{0,0,0.5}

\begin{document}

\title{Non-Gaussianities in Collider Metric Binning}%

\author{Andrew J.~Larkoski}%
\email{larkoski@aps.org}
\affiliation{American Physical Society, Hauppauge, New York 11788, USA}
\date{\today}%

\begin{abstract}
\noindent Metrics for rigorously defining a distance between two events have been used to study the properties of the dataspace manifold of particle collider physics.  The probability distribution of pairwise distances on this dataspace is unique with probability 1, and so this suggests a method to search for and identify new physics by the deviation of measurement from a null hypothesis prediction.  To quantify the deviation statistically, we directly calculate the probability distribution of the number of event pairs that land in the bin a fixed distance apart.  This distribution is not generically Gaussian and the ratio of the standard deviation to the mean entries in a bin scales inversely with the square-root of the number of events in the data ensemble.  If the dataspace manifold exhibits some enhanced symmetry, the number of entries is Gaussian, and further fluctuations about the mean scale away like the inverse of the number of events.  We define a robust measure of the non-Gaussianity of the bin-by-bin statistics of the distance distribution, and demonstrate in simulated data of jets from quantum chromodynamics sensitivity to the parton-to-hadron transition and that the manifold of events  enjoys enhanced symmetries as their energy increases.
\end{abstract}

\maketitle

\noindent Non-Gaussianities in the statistical fluctuations of multi-point correlation functions are evidence for underlying short-distance interactions.  This is perhaps most well-known in the case of temperature fluctuations in the cosmic microwave background radiation, where through observational measurement of the three-point correlation function or bispectrum \cite{Bennett:1996ce,Komatsu:2001wu,WMAP:2008lyn,WMAP:2010qai,Planck:2015zfm,Planck:2019kim,ParticleDataGroup:2024cfk}, parameters in predictions from models of cosmic inflation, e.g., Refs.~\cite{Maldacena:2002vr,Bartolo:2004if}, can be bounded or ruled out.

For the cosmic microwave background, the measurement or prediction of non-Gaussianities is straightforward, if technically challenging, because the space on which the data live is the two-dimensional celestial sphere.  Distances between different points on the sky are well-defined and unambiguous as simply the angle between them.  Additionally inspired by results of the two- and three dimensional Ising models \cite{Belavin:1984vu,Rychkov:2016mrc}, there has been recent interest in applying a study of non-Gaussianities to data in particle collider physics \cite{Chen:2022swd}.  Observing that physics at disparate scales factorizes from one another, the non-Gaussianity measure proposed in Ref.~\cite{Chen:2022swd} effectively isolates three-particle correlations from iterated two-particle correlations in each event individually, and then averages over events.

While such a measure of higher-point correlations is useful and informative, it is rather distinct from what is measured on the cosmic microwave background.  As close as possible analogously, we would like to measure non-Gaussianities on the ``universe'' of particle collider events, from multi-point correlations between entire events themselves.  This approach requires a robust definition of distances between events because the space on which particle collider data live is extremely high dimensional and varies event-by-event.  A number of such collider event metrics have recently been proposed \cite{Komiske:2019fks,Mullin:2019mmh,CrispimRomao:2020ejk,Cai:2020vzx,Larkoski:2020thc,Tsan:2021brw,Cai:2021hnn,Kitouni:2022qyr,Alipour-Fard:2023prj,Larkoski:2023qnv,Davis:2023lxq,Ba:2023hix,Craig:2024rlv,Cai:2024xnt,Gambhir:2024ndc,Larkoski:2025clo,Cai:2025fyw}.  Additionally, because collider event data is continuous-valued and drawn from some continuous distribution, there are general proofs that the probability distribution of pairwise distances between events, or the distribution of two-point event correlations,  is unique with unit probability \cite{boutin2004reconstructing}.

Our goal in this Letter will be to establish the finite statistical properties of the distribution of distances between events in some dataset.  This analysis will be completely general, but we will consider specific motivated examples from particle collider physics to illustrate our results.  We assume that our dataset consists of $n$ events that are i.i.d.~on coordinate space according to some distribution $p_E$.  In our collider physics context, this would be particle momentum phase space, and we denote individual points on this space corresponding to a complete event and the momentum of all of its constituent particles as $\vec x$.  We assume that the dataspace is also endowed with a metric $d(\cdot,\cdot)$ which returns the distance between two events.  As a metric, $d(\cdot,\cdot)$ satisfies the properties of identity of indiscernibles, positivity, symmetry, and the triangle inequality on almost all of the dataspace.

The distribution of pairwise distances $\epsilon$ between events on the dataspace is
\begin{align}\label{eq:pwdist}
p(\epsilon) = \int d\vec x\, d\vec x'\, p_E(\vec x)\, p_E(\vec x')\,\delta\left(
\epsilon - d(\vec x,\vec x')
\right)\,,
\end{align}
where $\delta(x)$ is the Dirac $\delta$-function.  On any finite dataset of $n$ events, we have to bin the distribution in $\epsilon$, with some finite width $\delta\epsilon > 0$.  The number of pairs of events $N_\epsilon$ that would be placed in the bin at distance $\epsilon$ is
\begin{align}
\hspace{-0.2cm}N_\epsilon = \!\!\!\!\sum_{1\leq i<j\leq n}\!\!\!\!\Theta(
\epsilon+\delta\epsilon - d(\vec x_i,\vec x_j))\Theta(
 d(\vec x_i,\vec x_j)-\epsilon
),
\end{align}
where $\Theta(x)$ is the Heaviside step function.  To avoid double-counting, we consider events as ordered in their label $i$.  As all events are i.i.d.~on phase space, the number of events $j$ that lie a distance of about $\epsilon$ from event $i$ at phase space point $\vec x_i$, with indices $i<j$, is Gaussian distributed as
\begin{align}
p(n_{\epsilon,i}|\vec x_i) = \frac{e^{-\frac{( n_{\epsilon,i}-\langle n_{\epsilon,i}\rangle)^2}{2\langle n_{\epsilon,i}\rangle}}}{\sqrt{2\pi\langle n_{\epsilon,i}\rangle}}\,.
\end{align}
The mean and variance is
\begin{align}
\langle n_{\epsilon,i}\rangle 
&= (n-i)\,\delta\epsilon\int d\vec x\, p_E(\vec x)\,\delta\left(\epsilon-d(\vec x_i,\vec x)\right)\,,
\end{align}
where we assume the bin width is much smaller than the location of the bin, $\delta\epsilon\ll \epsilon$, and the number of events with index larger than $i$ is large, $n-i\gg 1$. 

The distribution of the number of total pairs of events in the dataset that are about a distance $\epsilon$ from one another can then be expressed through convolution of all possible individual event pairings, where
\begin{align}
p(N_\epsilon) &= \int \prod_{1\leq i<n}\left[
dn_{\epsilon,i}\,d\vec x_i\,p(n_{\epsilon,i}|\vec x_i)\, p_E(\vec x_i)
\right]\\
&\hspace{3cm}\times\delta\left(
N_\epsilon -\sum_{1\leq i< n}n_{\epsilon,i}
\right)\nonumber\,.
\end{align}
We can construct the cumulant generating function of this distribution by taking the logarithm of the Laplace transform.  In the large number of events limit $n\to\infty$, we define
\begin{align}\label{eq:cumgenfct}
K_\epsilon(\tau) &\equiv \log\left[
\int dN_\epsilon\, e^{\tau N_\epsilon}\, p(N_\epsilon)
\right]\\
&=\int_0^n d\alpha \log\left[
\int d\vec x\, p_E(\vec x)\, e^{\alpha\, \delta\epsilon\,p(\epsilon|\vec x)\,\frac{\tau^2+2\tau}{2}
}
\right]\,,\nonumber
\end{align}
explicitly integrating over the Gaussian conditional probabilities.  The conditional probability of distance $\epsilon$ at point $\vec x$ is
\begin{align}
p(\epsilon|\vec x) \equiv \int d\vec x'\, p_E(\vec x')\,\delta\left(
\epsilon - d(\vec x,\vec x')
\right)\,.
\end{align}

While neither the distribution of number of pairs of events nor its cumulant generating function can be evaluated in general, we can systematically calculate its moments.  The mean number of pairs of events in the bin around distance $\epsilon$ is
\begin{align}
\langle N_\epsilon\rangle &= \left.\frac{d}{d\tau}K_\epsilon(\tau)\right|_{\tau=0} \!\!\!= \frac{n^2}{2}\delta\epsilon\!\int d\vec x\, p_E(\vec x)\,p(\epsilon|\vec x)\,,
\end{align}
as expected from the definition of the distribution of pairwise distances, Eq.~\ref{eq:pwdist}.  The variance of the number of pairs of events in this bin is 
\begin{align}\label{eq:varpair}
\sigma_\epsilon^2 &= \langle N_\epsilon^2\rangle - \langle N_\epsilon\rangle^2 = \left.\frac{d^2}{d\tau^2}K_\epsilon(\tau)\right|_{\tau=0}\\
&=\frac{4}{3n}\left(
\frac{n^4}{4}\,\delta\epsilon^2\!\!\int d\vec x\,p_E(\vec x)\,p(\epsilon|\vec x)^2-\langle N_\epsilon\rangle^2
\right)+\langle N_\epsilon\rangle
\nonumber\,.
\end{align}

The term in parentheses in Eq.~\ref{eq:varpair} involves honest three-point or three-event correlations, and so corresponds to non-Gaussianities in the distribution of the number of pairs of events in this bin.  Further, the contribution to the variance from such non-Gaussianities scales like $n^3$ in the large number of events $n$ limit, while the mean only scales like $n^2$.  As such, we generically expect that non-Gaussianities dominate the variance with fixed bin width and large number of events $n$.  In this the case, the ratio of the standard deviation to the mean scales like $n^{-1/2}$,
\begin{align}
\lim_{n\to\infty}\frac{\sigma_\epsilon}{\langle N_\epsilon\rangle}\propto n^{-1/2}\,.
\end{align}
Intriguingly, even though the number of pairs that contributes to the distance distribution is proportional to $n^2$, the distances are correlated and this correlation reduces the relative scaling of the standard deviation to be exactly as expected from Poissonian statistics of $n$ i.i.d.~events.

There may exist dataspace manifolds for which this $n^3$ contribution to the variance vanishes, when
\begin{align}
\frac{n^4}{4}\,\delta\epsilon^2\!\!\int d\vec x\,p_E(\vec x)\,p(\epsilon|\vec x)^2=\langle N_\epsilon\rangle^2\,.
\end{align}
This equality can be satisfied only if the probability $p(\epsilon|\vec x)$ is independent of phase space point $\vec x$.  In that case, this implies that the dataspace exhibits some enhanced symmetry at the distance scale $\epsilon$ so that the probability is unchanged by translation from point $\vec x$ to another point on the space.  If this relationship holds, the cumulant generating function in Eq.~\ref{eq:cumgenfct} can be evaluated in closed form as
\begin{align}
\hspace{-0.308cm}K_\epsilon(\tau) = \frac{n^2}{2}\delta\epsilon\,\frac{\tau^2+2\tau}{2}\!\!\int d\vec x\, p_E(\vec x)\,\delta(
\epsilon - d(\vec 0,\vec x)
),
\end{align}
where $\vec 0$ is the (arbitrary) origin on the space.  This cumulant generating function is quadratic, and so the distribution of number of pairs of events in this bin is Gaussian, with mean and variance equal.  Further, deviations from the mean scale away like $n^{-1}$,
\begin{align}
\lim_{n\to\infty}\frac{\sigma_\epsilon}{\langle N_\epsilon\rangle} = \frac{1}{\sqrt{\langle N_\epsilon\rangle}}\propto n^{-1}\,.
\end{align}

This observation suggests a definition of non-Gaussianities on the dataspace, as well as a practical procedure for establishing the presence of some symmetry in the data, exclusively from the scaling of finite statistical fluctuations.  We define the non-Gaussianity measure $\eta_\text{n-G}(\epsilon)$ at scale $\epsilon$ to be
\begin{align}\label{eq:nongausdef}
&\eta_\text{n-G}(\epsilon) \equiv \frac{\sigma_\epsilon^2 - \langle N_\epsilon\rangle}{\langle N_\epsilon\rangle} \\
&=\frac{2n\,\delta\epsilon}{3}\frac{\int d\vec x\,p_E(\vec x)\,p(\epsilon|\vec x)^2-\left(
\int d\vec x\, p_E(\vec x)\,p(\epsilon|\vec x)
\right)^2}{\int d\vec x\, p_E(\vec x)\,p(\epsilon|\vec x)}\nonumber\,.
\end{align}
This is non-negative and only vanishes if the fluctuations in the bin about distance $\epsilon$ are exactly Gaussian.

Before applying this analysis to collider physics events, it is insightful to study a simple example that can be solved completely analytically.  Our dataspace will be $n$ points uniformly distributed on the line of unit length and we take the Euclidean metric to define pairwise distances.  To determine the mean and variance of the number of pairs that lie in the bin about distance $\epsilon$, we need the the probability distribution $p(\epsilon|x)$, where
\begin{align}
p(\epsilon| x)&=\int_0^1dx'\,\delta\left(\epsilon - |x-x'|\right) \\
&= \Theta(x-\epsilon)+\Theta(1-\epsilon-x)\,.
\nonumber
\end{align}
The non-Gaussianity measure as a function of distance $\epsilon$ is
\begin{align}\label{eq:ngline}
\eta_\text{n-G}(\epsilon) =\frac{2n}{3}\,\delta\epsilon\left|
1-2\epsilon
\right|\min\left[
\frac{\epsilon}{1-\epsilon},1
\right]
\,.
\end{align}
Besides the point $\epsilon = 0$, this vanishes at $\epsilon = 1/2$, where the fluctuations of events in that bin are thus Gaussian.  In this bin, (almost) every point on the interval $x\in[0,1]$ has exactly one partner point that is a distance $\epsilon = 1/2$ away, while in bins at larger or smaller distances, points may have 0 or 1 partner points, or 1 or 2 partner points, respectively.

\begin{figure}[t!]
\begin{center}
\includegraphics[width=0.47\textwidth]{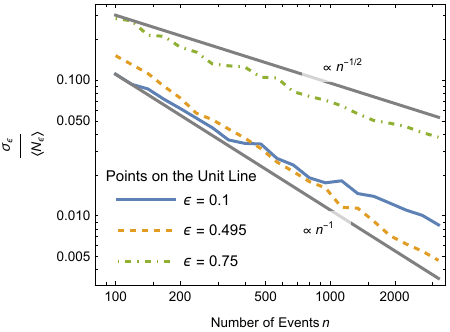}
\caption{\label{fig:linepts}
Plots of the ratio of the standard deviation $\sigma_\epsilon$ of the number of pairs of events in the bin about distance $\epsilon$ to the mean number of pairs $\langle N_\epsilon\rangle$, as a function of the total number of points $n$ sampled on the unit line.  
}
\end{center}
\end{figure}

In Fig.~\ref{fig:linepts}, we illustrate the scaling with number of points of the ratio of the standard deviation to the mean number of pairs of events in three different $\epsilon$ distance bins.  We consider two generic distances, $\epsilon = 0.1$ and $\epsilon = 0.75$, and near the symmetric point in the bin $\epsilon = 0.495$, and use bin width of $\delta\epsilon = 0.01$ for each.  The bin at $\epsilon = 0.75$ is observed to scale closely to $n^{-1/2}$, while the bin at $\epsilon = 0.495$ scales like $n^{-1}$, both as expected from the general analysis.  The bin at $\epsilon = 0.1$ scales like some intermediate power, but this is because that value of $\epsilon$ corresponds to a small non-Gaussianity.  From Eq.~\ref{eq:ngline}, we have $\eta_\text{n-G}(\epsilon = 0.1) = 0.059\,n\,\delta\epsilon$, so non-Gaussianities only dominate at sufficiently large number of points $n$ with fixed bin width $\delta\epsilon$.

For the application of this analysis to collider physics, we consider the space of jets, manifestations of quantum chromodynamics (QCD) that are collimated, high-energy streams of particles copiously produced at a hadron collider.  For calculational simplicity and speed when evaluating on (simulated) data, we use the $p=2$ Spectral Energy Mover's Distance (SEMD) metric \cite{Larkoski:2023qnv,Gambhir:2024ndc} which enjoys the properties of infrared and collinear safety \cite{Sterman:1977wj,Ellis:1996mzs}, and so can be calculated in the perturbation theory of QCD, and is also invariant to isometries of events.

We present a simple calculation of the scaling of the non-Gaussianities on the space of QCD jets, and then validate the predictions in simulated data.  We work to double logarithmic accuracy (DLA) at which dominant emissions are both soft and collinear to the initiating particle, the QCD coupling $\alpha_s$ is assumed to not run with energy scale, and the $p=2$ SEMD is the squared metric and is proportional to the sum of the squared masses of the pair of jets $A$ and $B$:
\begin{align}
d^{(\text{DLA})}(\vec x_A,\vec x_B)^2 = m_A^2+m_B^2\,.
\end{align}
To this accuracy, the distribution of the squared mass $m^2$ of a jet is dominated by an exponential Sudakov factor \cite{Catani:1992ua},
\begin{align}
p^{(\text{DLA})}(m^2)\sim \exp\left[
-\frac{\alpha_s C}{2\pi}\log^2\frac{m^2}{E^2}
\right]\,,
\end{align}
where $E$ is the jet energy, and $C$ is the quadratic Casimir of SU(3) color of the initiating particle.

From these observations, we predict the scaling of the non-Gaussianities of the space of jets, noting that every instance of the distribution $p_E(\vec x)$ in Eq.~\ref{eq:nongausdef} is associated with a Sudakov factor.  The numerator of that expression thus contains either 3 or 4 Sudakov factors, while the denominator only has 2.  To double logarithmic accuracy, the exponential dependence of the non-Gaussianities on the space of jets scales like
\begin{align}
\eta^{(\text{DLA})}_\text{n-G}(\epsilon)\sim \frac{2}{3}\,n\,\delta\epsilon\, \exp\left[
-\frac{\alpha_s C}{2\pi}\log^2\frac{\epsilon^2}{E^2}
\right]\,,
\end{align}
in the small distance $\epsilon\ll E$ limit.  Therefore, at a fixed distance $\epsilon$ the non-Gaussianities die away at an exponential rate as the jet energy $E$ increases.

This naive double logarithmic analysis will, however, cease being a relevant approximation when non-perturbative physics dominates.  We can estimate the inter-jet distance $\epsilon_\text{np}$ at and below which the jet is best described by a gas of hadrons.  Non-perturbative physics becomes relevant when the dominant transverse momentum to the jet axis is comparable to the QCD scale $\Lambda_\text{QCD}\sim 1$ GeV \cite{Dokshitzer:1995zt,Dokshitzer:1995qm}.  As the SEMD metric is closely related to the sum of the squared masses of the jets, the non-perturbative distance $\epsilon_\text{np}$ will scale like
\begin{align}
\epsilon_\text{np}\sim \sqrt{2\Lambda_\text{QCD} ER}\,,
\end{align}
where $R$ is the jet radius.  

\begin{figure}[t!]
\begin{center}
\includegraphics[width=0.47\textwidth]{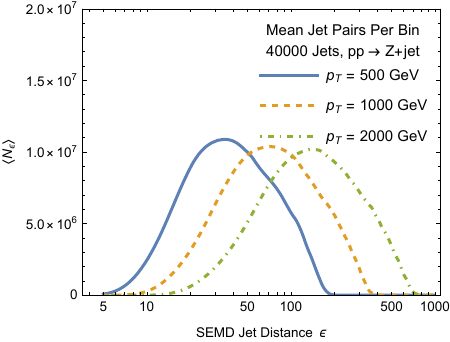}
\caption{\label{fig:means}
Plot of the mean number of pairs of jets $\langle N_\epsilon\rangle$ as a function of their SEMD metric distance $\epsilon$ apart.
}
\end{center}
\end{figure}

We generate $pp\to Z+$jet events at the 13 TeV Large Hadron Collider (LHC) at leading order with MadGraph v3.6.1 \cite{Alwall:2014hca}, and events are showered and hadronized with Pythia v8.306 \cite{Bierlich:2022pfr} with default settings.  We find anti-$k_T$ jets \cite{Cacciari:2008gp} with radius $R = 0.5$ with FastJet v3.4.0 \cite{Cacciari:2011ma} and record the most energetic jet in the events that lies in a transverse momentum bin of $p_\perp\in [500,550]$, $[1000,1100]$, or $[2000,2200]$ GeV.  Jets were reclustered into a maximum of 100 particles with the exclusive $k_T$ algorithm \cite{Ellis:1993tq,Catani:1993hr}.  We select $n=40000$ jets in each transverse momentum bin and calculate the full pairwise SEMD matrix.  We calculate the mean number of pairs of jets per bin $\langle N_\epsilon\rangle$ and the corresponding non-Gaussianity $\eta_\text{n-G}(\epsilon)$.  For both, we take 200 bins logarithmically distributed on $\epsilon \in [5,1000]$ GeV.

In Fig.~\ref{fig:means}, we plot the distribution of the number of pairs of jet events per bin, in the three transverse momenta ranges we consider.  These distributions have the characteristic bell shape on logarithimic axes, and are approximately just translated in $\epsilon$ from one another, as expected from the approximate scale invariance of QCD.  For these three samples, the approximate scale at which non-perturbative physics dominates is below $\epsilon_\text{np}\sim 22$, $32$, or $45$ GeV, respectively, each of which lies just to the left of the peaks of these distributions.

\begin{figure}[t!]
\begin{center}
\includegraphics[width=0.45\textwidth]{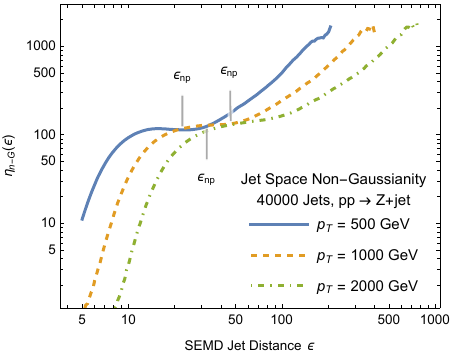}
\caption{\label{fig:nongauss}
Plot of the non-Gaussianity measure $\eta_\text{n-G}(\epsilon)$ on the simulated jet samples as a function of SEMD metric distance $\epsilon$.  The approximate locations of the distances at and below which non-perturbative physics dominates $\epsilon_\text{np}$ is identified on each curve.
}
\end{center}
\end{figure}

In Fig.~\ref{fig:nongauss}, we plot the non-Gaussianity of the distribution of these samples of jets on their respective dataspaces, as a function of distance $\epsilon$.  At intermediate distances, we see that the non-Gaussianity is reduced at higher jet energy, as expected from the simple double logarithmic analysis.  Very intriguingly, there is a plateau in all three datasets at the same value of non-Gaussianity. This plateau occurs around the distance $\epsilon_\text{np}$ above which the quasiparticles of jets are quarks and gluons, and below which hadrons are the natural degrees of freedom.  This feature may be indicative of a phase transition of QCD, as the functional inverse of the non-Gaussianity becomes non-analytic in this region.  The study of the statistics on the dataspace manifold could then be complementary to other techniques recently introduced for studying the parton-to-hadron transition, e.g., Refs.~\cite{Dreyer:2018nbf,Komiske:2022enw}.

In this Letter, we have introduced a measure of non-Gaussianities on the universe of particle collider events.  The analysis presented here connects to at least two programs in understanding the response and output of machine learning.  In particle physics specifically, there has been an effort to identify symmetries in data \cite{Krippendorf:2020gny,Tombs:2021wae,Lester:2021aks,Dillon:2021gag,Desai:2021wbb,Craven:2021ems,Birman:2022xzu,Forestano:2023fpj,Forestano:2023qcy,Calvo-Barles:2023num,Forestano:2023edq}, and the simple connection between symmetries and the scaling laws of fluctuations can provide an avenue for further insights.  More generally, scaling laws in machine learning \cite{ahmad1988scaling,cohn1990can,hestness2017deep,kaplan2020scaling,rosenfeld2019constructive,henighan2020scaling,rosenfeld2021predictability,Batson:2023ohn} have been observed in numerous contexts, and the dependence on manifold symmetries identified here may provide part of an underlying explanation.  We look forward to more connections and deeper understanding of the universe of particle collider events that can be provided through its metric space lens.

I thank Rikab Gambhir, Yoni Kahn, Gregor Kemper, Ian Moult, and Josef Schicho for comments.

\bibliography{refs}

\end{document}